\documentclass[useAMS,usenatbib, ]{mn2e}
\usepackage{aas_macros}
\usepackage{comment}
\usepackage{graphicx}
\usepackage{graphics}
\usepackage{epstopdf} 
\usepackage[usenames,dvipsnames]{xcolor}

\usepackage{amsmath}
\usepackage{url}
\usepackage{fixltx2e} 

\title[The East Cloud]{Major Substructure in the M31 Outer Halo: the East Cloud\thanks{This work was based on observations obtained with the MegaPrime/MegaCam, a joint project of CFHT and CEA/DAPNIA, at the Canada-France-Hawaii Telescope (CFHT) which is operated by the National Research Council (NRC) of Canada, the Institute National des Sciences de l'Univers of the Centre National de la Recherche Scientifique of France and the University of Hawaii.}}
\author[B. McMonigal et al.]
	{B. McMonigal,$^{1}$\thanks{E-mail: b.mcmonigal@physics.usyd.edu.au (BM)}, N. F. Bate$^{2}$, A. R. Conn$^{1}$, A. D. Mackey$^{3}$, G. F. Lewis$^{1}$, \newauthor M. J. Irwin$^{2}$, N. F. Martin$^{4, 5}$, A. W. McConnachie$^{6}$, A. M. N. Ferguson$^{7}$, \newauthor R. A. Ibata$^{4}$, A. P. Huxor$^{8}$\\
$^{1}$Sydney Institute for Astronomy, School of Physics, A28, University of Sydney, Sydney, NSW 2006, Australia\\
$^{2}$Institute of Astronomy, Madingley Road, University of Cambridge, CB3 0HA, UK\\
$^{3}$RSAA, The Australian National University, Mount Stromlo Observatory, Cotter Road, Weston Creek, ACT 2611, Australia\\
$^{4}$Observatoire Astronomique, Universite de Strasbourg, CNRS, F-67000 Strasbourg, France\\
$^{5}$Max-Planck-Institut f\"{u}r Astronomie, K\"{o}nigstuhl 17, D-69117 Heidelberg, Germany\\
$^{6}$NRC Herzberg Institute of Astrophysics, 5071 West Saanich Road, Victoria, British Columbia V9E 2E7, Canada\\
$^{7}$Institute for Astronomy, University of Edinburgh, Royal Observatory, Blackford Hill, Edinburgh, EH9 3HJ, UK\\
$^{8}$Astronomisches Rechen-Institut, Zentrum f\"{u}r Astronomie der Universit\"{a}t Heidelberg, M\"{o}nchhofstra{\ss}e 12-14, 69120 Heidelberg, \\Germany.\\}
\begin{document}

\date{Accepted --. Received --; in original form --}

\pagerange{\pageref{firstpage}--\pageref{lastpage}} \pubyear{2014}

\maketitle

\label{firstpage}

\begin{abstract}
We present the first detailed analysis of the East Cloud, a highly disrupted diffuse stellar substructure in the outer halo of M31. 
The core of the substructure lies at a projected distance of $\sim100$~kpc from the centre of M31 in the outer halo, with possible extensions reaching right into the inner halo. 
Using Pan-Andromeda Archaeological Survey photometry of red giant branch stars, we measure the distance, metallicity and brightness of the cloud.
Using Hubble Space Telescope data, we independently measure the distance and metallicity to the two globular clusters coincident with the East Cloud core, PA-57 and PA-58, and find their distances to be consistent with the cloud. Four further globular clusters coincident with the substructure extensions are identified as potentially associated. 
Combining the analyses, we determine a distance to the cloud of $814^{+20}_{-9}$~kpc, a metallicity of $[Fe/H] = -1.2\pm0.1$, and a brightness of $M_V = -10.7\pm0.4$~mag. Even allowing for the inclusion of the potential extensions, this accounts for less than $20$ per cent of the progenitor luminosity implied by the luminosity-metallicity relation. 
Using the updated techniques developed for this analysis, we also refine our estimates of the distance and brightness of the South-West Cloud, a separate substructure analyzed in the previous work in this series.
\end{abstract}


\section{Introduction}

The study of galactic formation and evolution through the identification of fossil remnants, also known as galactic archaeology, is critical to building our understanding of the universe. A key part of this, according to standard $\Lambda$CDM theory, is hierarchical structure formation, the process of accretion by which galaxies merge and grow to form ever larger structures. The long dynamical timescales in the outer regions of galactic halos mean that they act as historical records of their long accretion history.
Major mergers, violent events which erase the visible history of the galaxy, are key processes to galactic formation. However, by studying the many smaller accretions still visible in the outer halo, we can piece together the history of the Local Group. This provides data critical to modelling structure formation, thus helping to gain a greater understanding of the underlying processes.

The Andromeda galaxy (M31), our nearest large neighbour, is a reasonable analogue for our own galaxy, and offers a panoramic view unavailable in the Milky Way. A tremendous amount of effort has already gone into the analysis of M31.  Extensive deep optical surveys such as the Pan-Andromeda Archaeological Survey (PAndAS; \citealt{McConnachie2009}) and the Spectroscopic and Photometric Landscape of Andromeda's Stellar Halo (SPLASH) survey \citep{Gilbert2012} have enabled the analysis of the various components which make up the galactic halo in unrivalled detail, including the bulk properties of the stellar halo itself (e.g. \citealt{Ibata2014}). Many new dwarf galaxies have been discovered (e.g. \citealt{Martin2013}), bringing the current total for M31 dwarf spheroidals up to 33, and the large homogeneous data coverage has enabled consistent comparison of their properties (e.g. \citealt{McConnachie2012}; Martin et al. in prep). 

The globular cluster (GC) population consists of well over 500 confirmed globular clusters (see the Revised Bologna Catalogue, \citealt{Galleti2004}), including almost a hundred outer halo objects (\citealt{Veljanoski2013} and references therein; \citealt{Huxor2014}). The outer halo GCs are preferentially spatially co-incident with diffuse stellar substructures, with a chance alignment of this significance of less than 1 per cent (\citealt{Mackey2010}; \citealt{Veljanoski2014}). 

Studies of diffuse stellar substructures in M31 have tended to focus on inner halo features, such as the north eastern shelf, the northern spur, the G1 clump, and the giant stellar stream (e.g. \citealt{Ferguson2005}; \citealt{Faria2007}; \citealt{Richardson2008}; \citealt{Bernard2015}). 
However, key properties of the halo such as the bulk rotation of the GC population \citep{Veljanoski2014}, or the plane of dwarf galaxies (\citealt{Ibata2013}; \citealt{Conn2013}), which itself exhibits consistent rotation, could only be uncovered by doing a complete analysis of the halo. Thus it is critical that we complete the analysis of the major substructures in the outer halo.

In \citet{Bate2014}, hereafter Paper I, 
we presented the first detailed analysis of a diffuse stellar substructure in the outer halo of M31, the South-West Cloud (SWC), determining its progenitor to be amongst M31's brightest dwarf galaxies prior to disruption. We concluded that at least one of the three GCs co-incident with the SWC is associated with the SWC \citep{Gilbert2012}, with a second association confirmed by a follow-up in \citet{Mackey2014}. Here we present the results of an analogous analysis of a similarly significant substructure located on the opposite side of M31 (in projection), dubbed the East Cloud \citep{Lewis2013}. We also apply improved and new techniques to reanalyze the properties of the SWC throughout.

In Section \ref{sec:data} we discuss the PAndAS data and contamination model used throughout this paper. Section \ref{sec:density} contains contamination-corrected stellar density maps of the regions around the SWC and East Cloud, while also showing the locations of the nearby globular clusters. 
In Section \ref{sec:distances} we use a Bayesian technique to measure the magnitude of the tip of the red giant branch (TRGB), and so estimate a distance to the East Cloud. The TRGB technique simultaneously fits for metallicity, and is described in full in a companion paper (Conn et al submitted). 
In Section \ref{sec:brightness} we use two methods for estimating the present day luminosity of both the East Cloud and the SWC. The `Dwarf Method' is an improved version of the technique we used in Paper I, which is calibrated against the many dwarf galaxies present in the PAndAS data. 
For the new `Population Method', we introduce `flux chains' developed in a companion paper (Martin et al in prep), spatially resolved forward modelled color-magnitude diagram fits to the PAndAS data for a collection of single stellar populations and a contamination population.

In Section \ref{sec:feh} we measure stellar metallicities by comparing them to theoretical isochrones for ancient red giant branch stars at the distance of M31. This metallicity is compared with the estimate from the TRGB method for consistency. 
Using HST data, we present the detailed analysis of the globular clusters in Section \ref{sec:gcs}. By fitting templates to the two globular clusters coincident with the largest East Cloud over-density, we are able to estimate their distances and metallicities. 
Finally, we discuss the implications of these measurements and conclude in Section \ref{sec:discussion}.

Throughout this paper, we assume the distance modulus to M31 to be $(m-M)_0 = 24.46 \pm 0.05$, or $779^{+19}_{-18}$ kpc \citep{Conn2012}.

\section{PAndAS Data}\label{sec:data}

\begin{figure*} 
  \includegraphics[width=130mm]{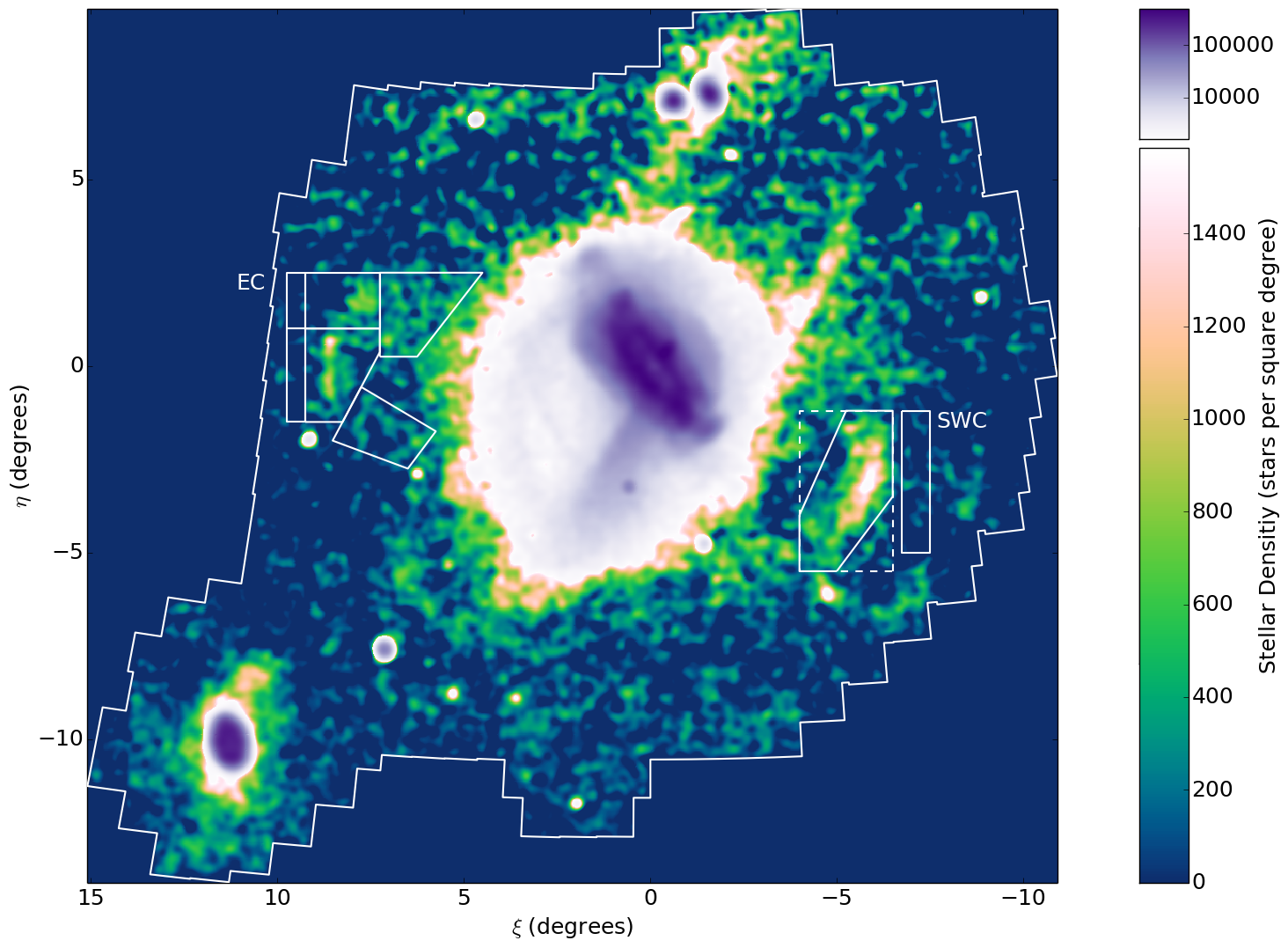}
  \caption{Stellar density map of PAndAS stars with de-reddened colours and magnitudes consistent with metal-poor red giant branch populations at the distance of M31. Pixels are $0.025\degr \times 0.025\degr$, and the map has been smoothed using a Gaussian with a dispersion of $\sigma = 0.1\degr$. It is displayed in tangent-plane projection centred on M31, with scaling chosen to highlight the East Cloud and South-West Cloud. This map has been foreground contamination-subtracted, as in Paper I (see \citealt{Martin2013} for full details). This and all subsequent colour maps in this paper were generated using the `cubehelix' scheme \citep{Green2011}.}
  \label{fig:map_full}
\end{figure*}

As discussed in Paper I, the photometry of M31 used in this paper was obtained as part of the PAndAS Large Program on the 3.6-metre Canada-France-Hawaii Telescope. Using the $0.96\times0.94$ square degree field of view MegaPrime camera, $g$ and $i$ band imaging was taken of M31 and M33, out to distances of approximately 150 kpc and 50 kpc respectively, with a total area of more than $\sim390$~deg$^2$.

A description of the PAndAS survey can be found in \citet{McConnachie2009}, with full details of the data reduction and a public release of the data in forthcoming publications (\citealt{Ibata2014}; McConnachie et al., in preparation). To summarise, all observations were taken in generally excellent seeing conditions ($\la0.8\arcsec$), with a mean seeing of $0.67\arcsec$ in $g$-band and $0.60\arcsec$ in $i$-band. The median depth is $g = 26.0$~mag and $i=24.8$~mag ($5\sigma$). 

The images were pre-processed by CFHT staff using their \textit{Elixir} pipeline, before being processed using a version of the CASU photometry pipeline \citep{Irwin2001} adapted for CFHT/MegaPrime observations. The resulting catalogue contains a single photometrically-calibrated $g$, $i$ entry for each detected object (see Paper I for more detail).

For this work, we use all objects in the final catalogue that have been reliably classified as stars in both bands (aperture photometry classifications of -1 or -2 in both $g$ and $i$, which corresponds to point sources up to $2\sigma$ from the stellar locus). The CFHT instrumental magnitudes $g$ and $i$ are converted to de-reddened magnitudes $g_0$ and $i_0$ on a source-by-source basis, using the following relationships from \citet*{Schlegel1998}: $g_0 = g - 3.793E(B-V)$ and $i_0 = i - 2.086E(B-V)$. 

Despite every effort to systematically cover the PAndAS survey region, holes do occur at the location of bright saturated stars, chip gaps, and a few failed CCDs. These holes have been filled with fake stars by duplicating information from nearby regions (for details, see \citealt{Ibata2014}). The fake entries are utilised to construct smooth stellar density maps, but are otherwise excluded from our calculations.

\subsection{Contamination model}
\label{sub:contamination}
Three sources of contamination in the PAndAS dataset can affect our study of substructure in the M31 outer halo: foreground Milky Way stars, unresolved compact background galaxies, and M31 halo stars. We account for this contamination using a model developed empirically from the PAndAS data in \citet{Martin2013}. In this model, the density of contaminants $\Sigma$ at a given location $(\xi, \eta)$ and a given colour and magnitude $(g_0-i_0, i_0)$ is given by a three component exponential:

\begin{equation}
\Sigma_{(g_0-i_0,i_0)}(\xi, \eta) = \rm{exp}(\alpha_{(g_0-i_0,i_0)}\xi + \beta_{(g_0-i_0,i_0)}\eta +\gamma_{(g_0-i_0,i_0)}).
\end{equation}

The coordinates $(\xi,\eta)$ in this model, and throughout this paper, are a tangent-plane projection centred on M31. The contamination model is defined over the colour and magnitude ranges $0.2\leq (g_0-i_0)\leq3.0$ and $20 \leq i_0 \leq 24$. This model allows us to generate a colour-magnitude diagram (CMD) for contamination at any location in the PAndAS footprint. From these contamination CMDs, we can generate contamination luminosity functions and stellar densities for any region of the PAndAS survey. For full details, see \citet{Martin2013}.

\section{Stellar Density Maps}\label{sec:density}

\begin{figure*} 
  \includegraphics[width=130mm]{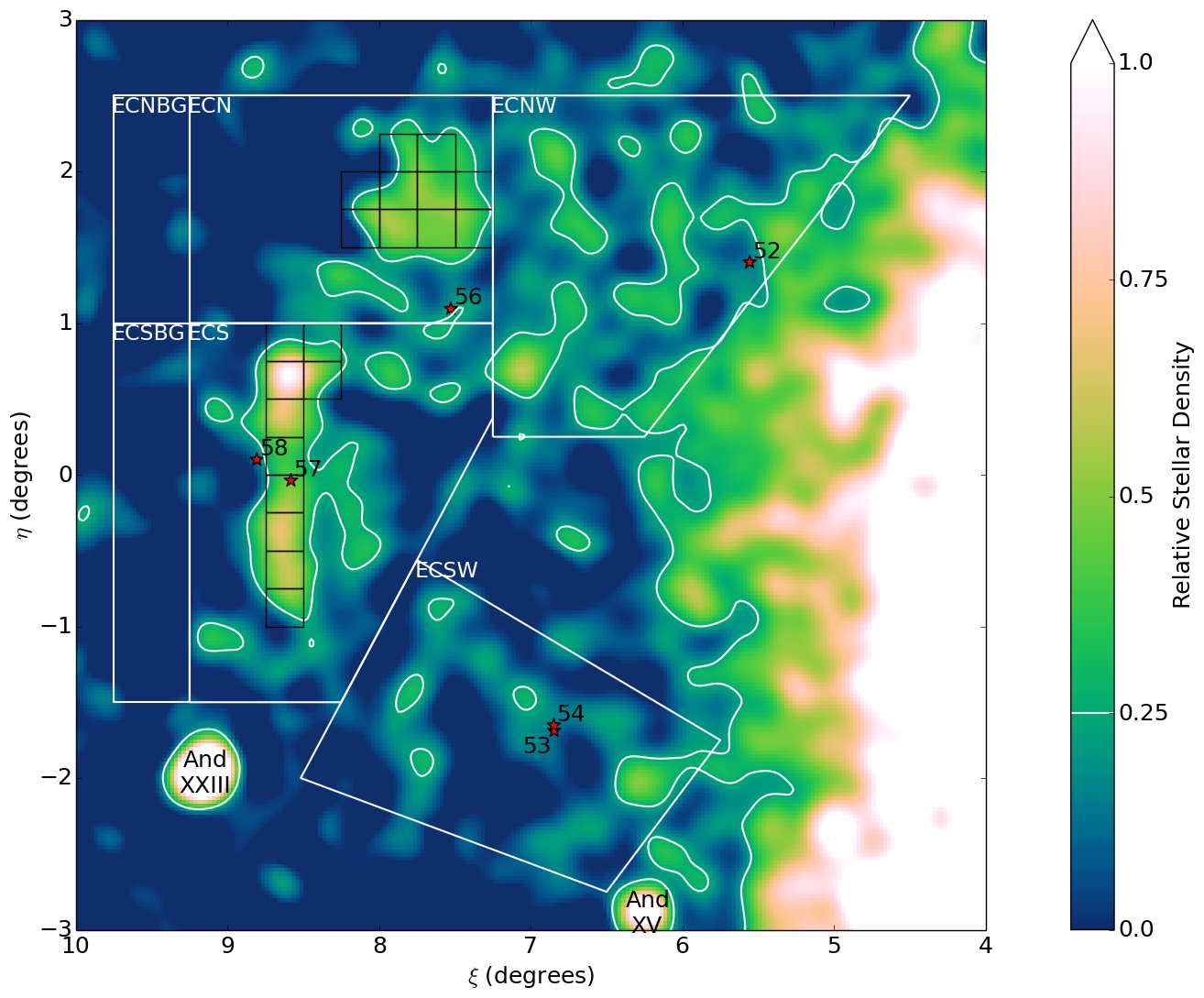}
  \caption{Map of East Cloud contamination-subtracted relative stellar density. Pixels are $0.025\degr \times 0.025\degr$, and have been smoothed with a Gaussian with $0.10\degr$ dispersion. The colour gradient is scaled relative to the peak in the main body of the East Cloud subfields (which lies in ECS), with contours at a relative density of 0.25 (white). The positions of known globular clusters in the region are marked with red stars. The flux chain pixels used for surface brightness estimation have been outlined in black (see Section \ref{sec:chains}). The white boundaries mark four separate EC fields, and the two background fields used for the primary EC fields. The small bright overdensities to the South-East and South-West are the dwarf galaxies AndXXIII and AndXV respectively.}
  \label{fig:map_ec}
\end{figure*}

In Figure \ref{fig:map_full} we show a contamination-subtracted stellar density map in tangent-plane projection centred on M31 for the entire PAndAS region. Stars in this map were chosen to be consistent with metal-poor red giant branch populations at the distance of M31, using a colour-magnitude box defined by the following $(g_0-i_0,i_0)$ vertices: $(0.4,23.5)$, $(0.7, 20.9)$, $(2.3, 20.9)$, $(1.6, 23.5)$. This captures the approximate metallicity range $-2.5 < [Fe/H] < -1.0$. The map is plotted on $0.025\degr \times 0.025\degr$ pixels, and smoothed with a Gaussian with dispersion $\sigma = 0.1\degr$.

\begin{figure}
  \includegraphics[width=80mm]{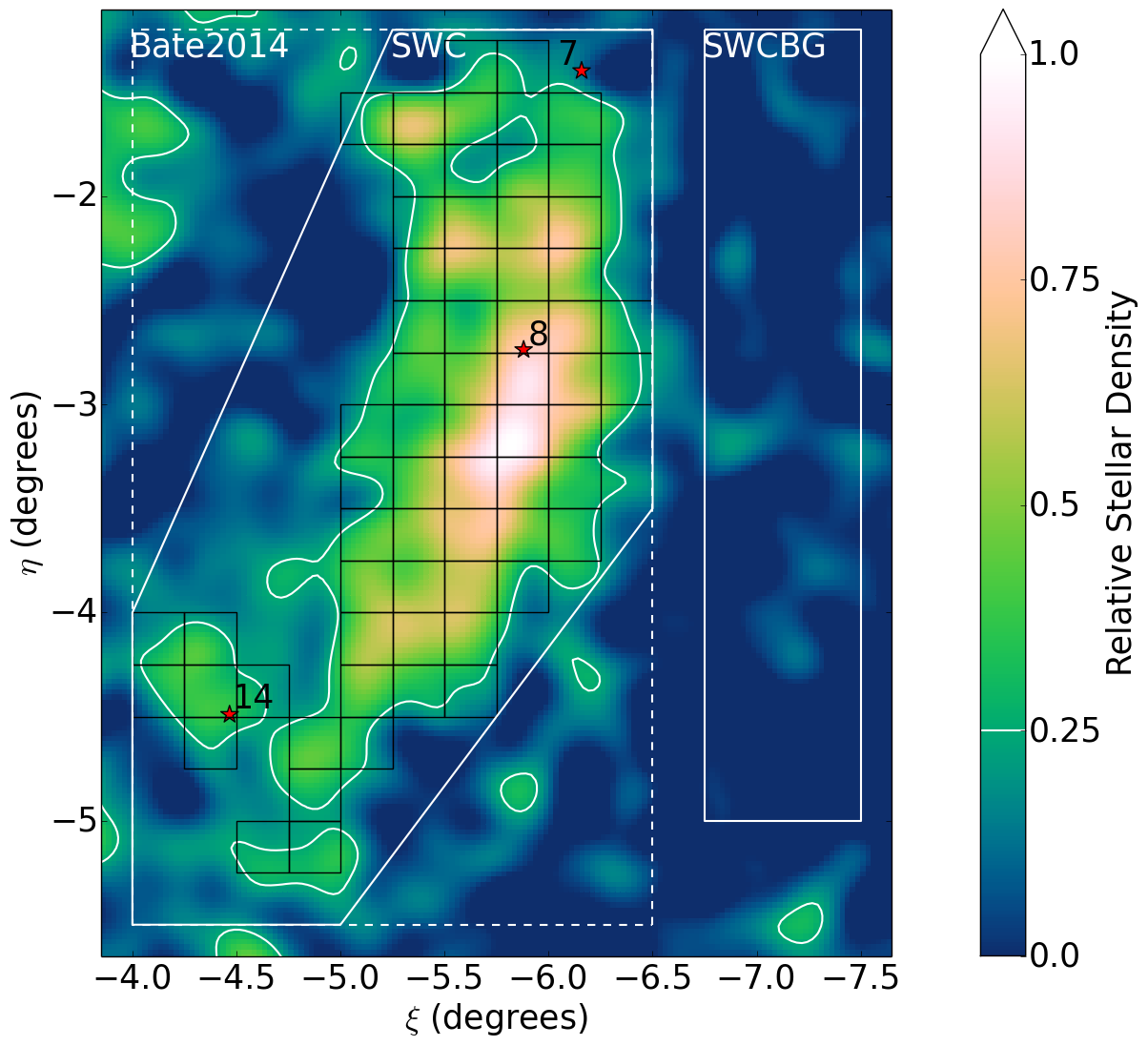}
  \caption{Map of South-West Cloud contamination-subtracted relative stellar density. Pixels are $0.025\degr \times 0.025\degr$, and have been smoothed with a Gaussian with $0.10\degr$ dispersion. The colour gradient is scaled relative to the peak in the main body of the SWC, with contours at a relative density of 0.25 (white). The positions of known globular clusters in the region are marked with red stars. The flux chain pixels used for surface brightness estimation have been outlined in black (see Section \ref{sec:chains}). The white boundaries mark the original SWC field used by \citet{Bate2014} (dashed), the tighter SWC field used by this work (solid), and the background field for the SWC.}
  \label{fig:map_swc}
\end{figure}

The white boundaries mark the target fields, and where relevant their corresponding contamination fields. The focus of this paper, the East Cloud, is marked by the fields on the left of M31. Since Paper I, we have improved our techniques and developed new methods for brightness and distance estimation, so for consistency we also reanalyze the South-West Cloud, marked by the fields on the right of M31. These fields and their various subfields are shown in more detail in Figures \ref{fig:map_ec} and \ref{fig:map_swc}.

The South-West Cloud is made up of a single large cloud, with what looks like an extension to the South-East, as was discussed in Paper I. The East Cloud is even more disrupted, with a dominant cloud in the Southern field ECS, with a detached cloud to the North in ECN. The South-West and North-West show potential extensions in fields ECSW and ECNW respectively, forming a single large arc running through the four East-Cloud fields.

For this paper we focus on the main populations marked ECS and ECN, as the extensions are difficult to separate from the rest of the M31 population due to both their proximity to M31 and their very low surface brightness.
The remaining fields ECSBG, ECNBG, and SWCBG are used for contamination estimation for one of the brightness estimation methods. These fields were chosen for their proximity to their corresponding target fields ECS, ECN, and SWC, while avoiding obvious substructure, and giving a good representation of the full extent of the Galactic latitude the target fields encompass (represented by the $\eta$ coordinate).

There are six globular clusters in the vicinity of the East Cloud \citep{Huxor2014}, marked in Figure \ref{fig:map_ec}. All six appear to lie along the arc made by the East Cloud substructure, and the two globular clusters PAndAS-57 and PAndAS-58 lie very close in projection to the densest part of the substructure.

While stellar membership for dwarf galaxies is typically modelled by radial density profile, the structures we are interested in are so disrupted that this is not feasible. Instead, we use stellar density relative to the peak density in the target field.

For the purposes of this paper, we use the relative stellar density maps smoothed with the $0.10\degr$ dispersion Gaussian kernel as a method for determining membership probability. This dispersion was chosen to give a clear view of the connected region around each target. Stars are assigned a probability $p$ for membership in a target equal to the stellar density in the pixel in which they fall. This probability is scaled relative to the peak in the target field, such that a star lying in the peak density pixel has a membership probability of $p=1$.

Thus the relative stellar density maps for the East Cloud and the South-West Cloud, shown in Figures \ref{fig:map_ec} and \ref{fig:map_swc}, can be used as membership probability maps, however, these membership probabilities should only be taken seriously within fields ECS, ECN, and SWC. These maps are primarily used for distance estimation, but are also used to establish membership cutoffs for brightness and metallicity estimates. The cutoff used is a scaled stellar density of 0.25 of the target maximum (marked by a contour in the Figures).

\section{Distances}\label{sec:distances}

Here we estimate the distance to the East Cloud utilizing the entirety of the combined fields ECS and ECN.
For this estimate, we employ the Tip of the Red Giant Branch (TRGB) standard candle, determined by a model fitting technique as introduced in Conn et al (submitted). 
In summary, the method fits a three dimensional surface to the colour-magnitude diagram (CMD) of the combined stellar population, modelling the stellar density at a given magnitude and colour as the sum of a red giant branch component and a contamination component. 

The red giant branch component is modelled as an isochrone grid, using theoretical isochrones from the Dartmouth Stellar Evolution Database \citep{Dotter2008}. The metallicity range spanned by the grid is weighted by a Gaussian profile, such that isochrones at the centre of the Gaussian receive the highest weighting. We equate the centre of the Gaussian with the best-fit metallicity of the structure's red giant branch whilst the width of the Gaussian we shall refer to as the RGB width, which is distinct from the metallicity spread as calculated in Section \ref{sec:feh}. The isochrones are shifted along the magnitude axis to find the best fit location of the tip of the red giant branch. 

We fix the age of the isochrone grid at $13$ Gyr for our East Cloud (and South West Cloud) measurements as is the case throughout this contribution. The contamination component is a Hess diagram generated for the location of the East Cloud within the PAndAS survey area, as outlined in \citet{Martin2013}. The ratio of the RGB component to the contamination component is then an additional fitted parameter. The method is essentially a two-dimensional extension of the Luminosity Function fitting algorithm introduced in \citet{Conn2011} and \citet{Conn2012}. 

In order to model the photometric uncertainty in the PAndAS data, we convolve the model CMD generated as described above with a two-dimensional Gaussian kernel using an approximated uncertainty in both $i$ and $g$ bands of $0.015$ magnitudes. Whilst the uncertainty in our fitted magnitude range varies from approximately $0.005$ to $0.035$, the $0.015$ magnitudes used is a good approximation to the expected uncertainty within the region we expect to find the TRGB. With regard to blending, we have found that it is only an issue where the stellar density is very high, such as in the centres of some of M31's larger dwarf spheroidal satellites. Overcrowding of this kind is not an issue in any of the structures presented in this contribution. Likewise, data incompleteness is not an issue for the fitted magnitude range, with the survey reaching a depth of some $2$ to $3$ full magnitudes deeper than our cutoff.

\begin{figure}
  \includegraphics[width=80mm, angle=0]{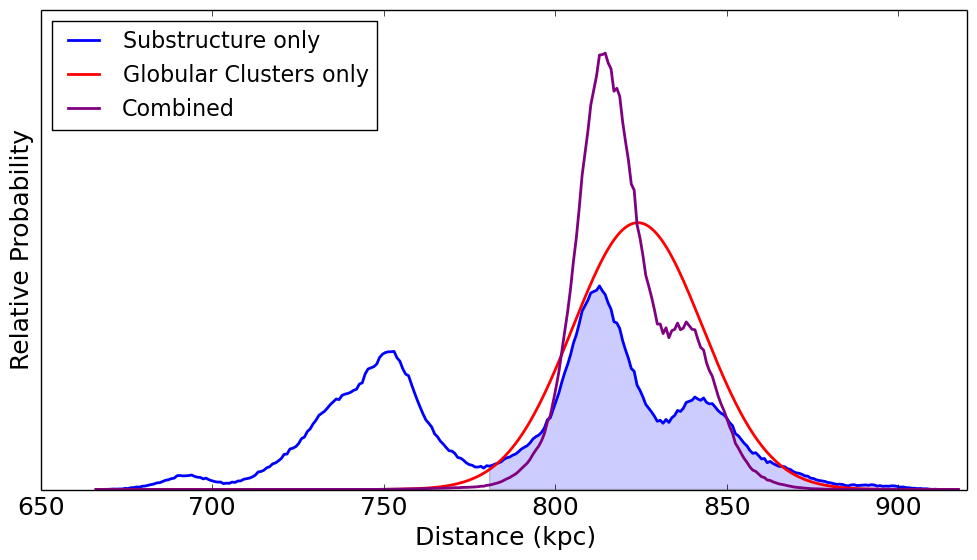}
  \caption{Probability distributions for the distance to the East Cloud. The probability distribution as determined from the combined fields ECS and ECN using the Tip of the Red Giant Branch standard candle is plotted in blue. The distribution determined from the horizontal branch of the globular clusters PA-57 and PA-58 is plotted in red (see Section \ref{sec:gcs}). The purple distribution is the multiplication of these two distributions, yielding our best-fit distance measurement to the East Cloud of $814^{+20}_{-9}$ kpc. The shaded region in blue indicates the region taken for a more restrictive measurement, guided by the combined fit.}
  \label{distance_pdf}
\end{figure}

Due to the very low contrast of the East Cloud (a contamination in excess of $98$ per cent is determined by our TRGB fitting algorithm), we obtain a very broad, multiple peaked probability distribution in the TRGB, indicating that the algorithm is heavily influenced by a small number of stars in the region of the tip. This probability distribution appears in Figure \ref{distance_pdf} in blue. 

The distance measurement obtained for the globular clusters PA-57 and PA-58 in Section \ref{sec:gcs} is represented by the red Gaussian. Since deeper photometry was available for this measurement, it was possible to utilize the horizontal branch as a standard candle -- which is much better populated than the region of the TRGB, hence the tighter distance constraints. Nevertheless, in multiplying the two distributions together (the purple distribution in Figure \ref{distance_pdf}), we are able to provide a further constraint on the distance, assuming the association of the globular clusters. From this distribution we find a best fit distance to the structure of $814^{+20}_{-9}$ kpc. 

\begin{figure}
  \includegraphics[width=80mm, angle=0]{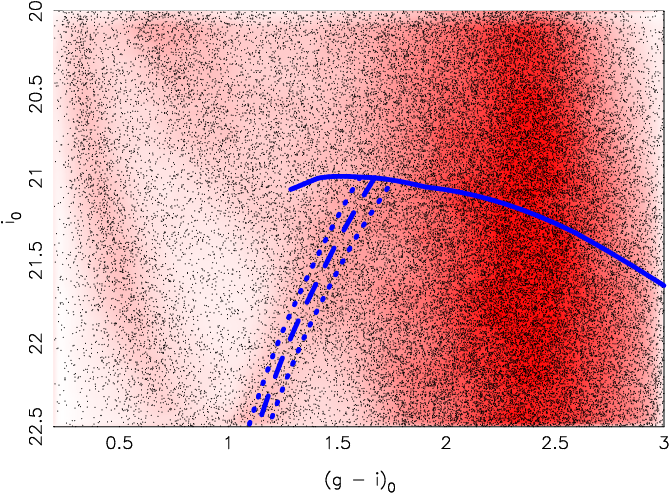}
  \caption{Best fit model generated by the TRGB algorithm for the East Cloud. 
  The model height (density) as a function of magnitude $i_0$ and color $(g_0 - i_0)$ is indicated by the shade of red at that location. The observed stars are marked by black dots. An isochrone representing the best-fit central metallicity of the data is shown as a blue dashed line. The blue dotted lines on either side are representative of the RGB width - they do not represent the uncertainty in the best-fit metallicity value. The solid blue line denotes the magnitude of the TRGB as a function of color.}
  \label{fig:distance_ec}
\end{figure}

Additionally, we obtain a best fit metallicity of $-1.4^{+0.2}_{-0.1}$ dex and a best fit RGB width of $0.3 \pm 0.1$ dex. Restricting the distance probability distribution to the broad peak consistent with the PA-57/PA-58 measurement (the blue shaded region in Figure \ref{distance_pdf}) has little effect except to reduce the uncertainty, yielding a best fit metallicity of $-1.4 \pm 0.1$ dex whilst the best fit RGB width is unchanged. The best fit model produced by the TRGB algorithm is shown in Figure \ref{fig:distance_ec}. The distance of the East Cloud from the centre of M31 is estimated at $111^{+13}_{-1}$ kpc, or alternatively $119^{+14}_{-4}$ kpc, if the restricted fitted range is taken. 

\begin{figure}
  \includegraphics[width=80mm, angle=0]{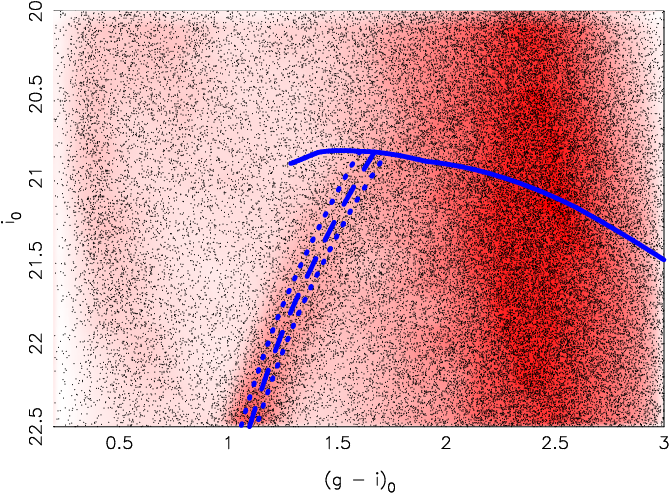}
  \caption{Best fit model generated by the TRGB algorithm for the SWC. 
  The model height (density) as a function of magnitude $i_0$ and color $(g_0 - i_0)$ is indicated by the shade of red at that location. The observed stars are marked by black dots. An isochrone representing the best-fit central metallicity of the data is shown as a blue dashed line. The blue dotted lines on either side are representative of the RGB width - they do not represent the uncertainty in the best-fit metallicity value. The solid blue line denotes the magnitude of the TRGB as a function of color.}
  \label{fig:distance_swc}
\end{figure}

In addition to the East Cloud, we also apply the new method outlined in Conn et al (submitted) 
to the South West Cloud, with the best fit model shown in Figure \ref{fig:distance_swc}. Via this method, we determine a distance to the cloud as a whole of $757^{+9}_{-30}$ kpc. This is consistent with the results recently published in \citet{Bate2014}, where a TRGB distance of $793^{+45}_{-45}$ kpc was estimated. This new distance corresponds to a distance from the centre of M31 of $85^{+7}_{-1}$ kpc. Additionally, we determine a metallicity of $-1.4 \pm 0.1$ dex which is in keeping with the $-1.3 \pm 0.1$ dex estimated by \citet{Bate2014}. As for the East Cloud, we find a relatively narrow RGB width, with a best fit value of $0.3 \pm 0.1$ dex proposed by our algorithm.

\section{Brightness}\label{sec:brightness}

Here we present brightness estimates for the main substructures of the East Cloud in ECS and ECN. We also analyse the SWC, both as a consistency check, and to update our estimate from Paper I, using the techniques developed for this analysis.
Two different methods are used to reconstruct the brightness of these substructures. The first method, which we call `the Dwarf Method', is an improvement upon the method used in Paper I, and uses the PAndAS data directly. The second method, which we call `the Population Method', uses spatially resolved forward modelled single stellar population fits to the PAndAS data, which we call `flux chains', detailed in Section \ref{sec:chains}. We discuss the methods in detail in Section \ref{sec:mags}, and present our total brightness estimates. Surface brightness estimates are discussed in Section \ref{sec:surf_mags}.

\subsection{Flux Chains}\label{sec:chains}

In a companion paper (Martin et al., in prep), we perform the forward modelling of the PAndAS CMDs for every $15\times15$ arcmin$^2$ pixel in the survey footprint. For a given pixel, the CMD is modelled as the contamination CMD expected for this location, as defined by \citet{Martin2013}, combined with the sum of 23 simple stellar populations. These stellar populations are at the distance of M31 ($(m-M)_0 = 24.46$), with an age of 13 Gyr, and with metallicity [Fe/H] that varies between $-2.3$ and $-0.1$ in $0.1$ dex increments. These stellar populations are based on the \textsc{Parsec} library isochrones and luminosity functions for a Kroupa (2001) initial mass function (IMF). 

The number of stars per component of the model (the contamination and the 23 simple stellar populations) within the M31 RGB region is inferred through a Markov Chain Monte Carlo approach by enforcing that they are strictly positive, that the number of stars in the contamination component follows a Poissonian prior centred around the number of stars expected in the \citet{Martin2013} model, and that the total number of stars in the pixel has a Poissonian prior centred on the observed number of stars in the CMD selection box for this pixel.

In an additional step, for each simple stellar population, the resulting numbers of stars within the CMD selection box are converted into surface brightnesses in the $g$ and $i$ bands through the determination of the fraction of flux contained in the selection box for the given stellar population (see Figure 1 of \citealt{Martin2013} for the location of this selection box). A given chain produced by the algorithm therefore contains a sampling of the 24-parameter probability density function (PDF) of surface brightnesses for the 23 simple stellar populations as well as the contamination component of the model, which we consider as a nuisance parameter for the current analysis.

\begin{figure}
  \includegraphics[width=80mm]{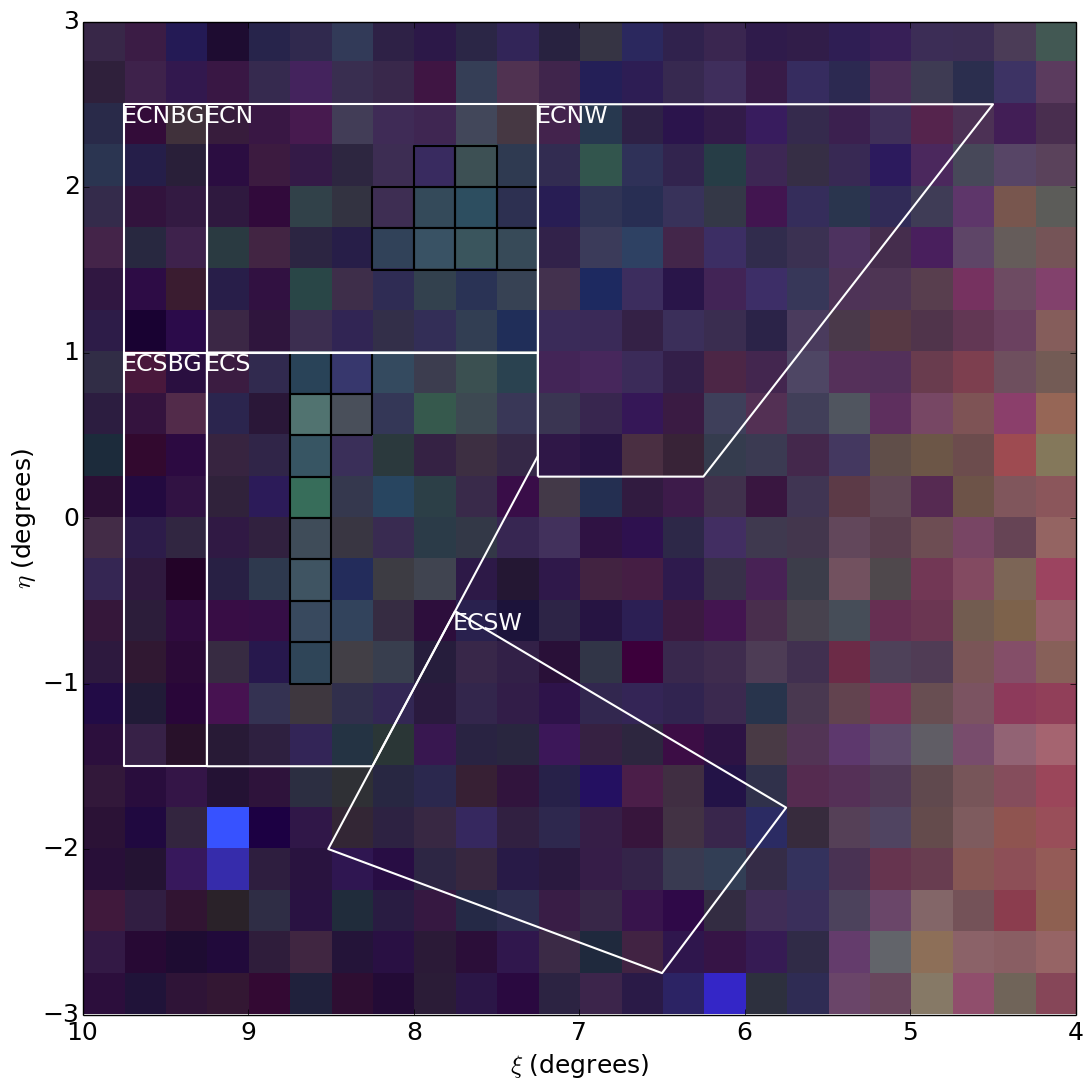}
  \caption{Map of the East Cloud flux based on the chains discussed in Section \ref{sec:chains}. Pixels are $0.25\degr \times 0.25\degr$. The populations are split into three metallicity ranges, $-2.3 \leq [Fe/H] \leq -1.5$ (low/blue), $-1.4 \leq [Fe/H] \leq -1.0$ (intermediate/green), and $-0.9 \leq [Fe/H] \leq -0.2$ (high/red). The flux chain pixels used for surface brightness estimation have been outlined in black. The white boundaries are as in Figure \ref{fig:map_ec}. The bright blue overdensities to the South-East and South-West are the dwarf galaxies AndXXIII and AndXV respectively.}
  \label{fig:map_ec_flux}
\end{figure}

\begin{figure}
  \includegraphics[width=80mm]{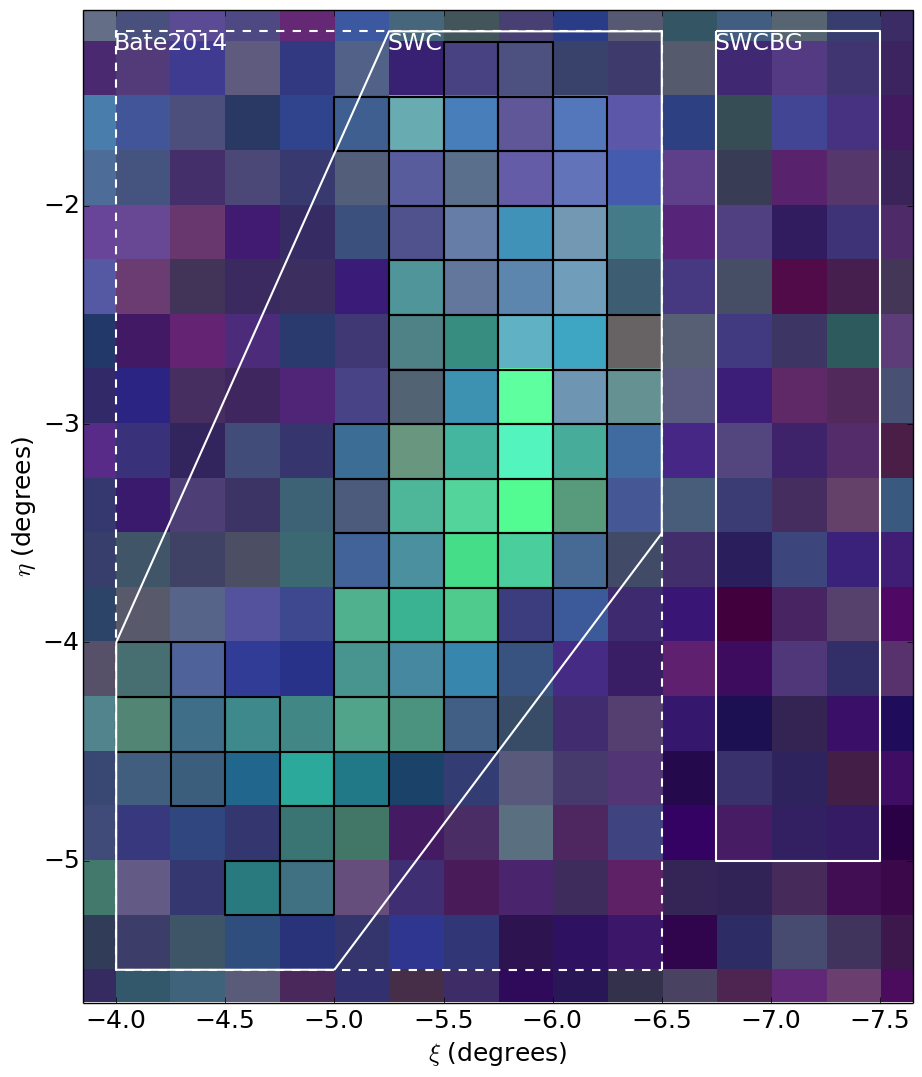}
  \caption{Map of the South-West Cloud flux based on the chains discussed in Section \ref{sec:chains}. Pixels are $0.25\degr \times 0.25\degr$. The populations are split into three metallicity ranges, $-2.3 \leq [Fe/H] \leq -1.5$ (low/blue), $-1.4 \leq [Fe/H] \leq -1.0$ (intermediate/green), and $-0.9 \leq [Fe/H] \leq -0.2$ (high/red). The flux chain pixels used for surface brightness estimation have been outlined in black. The white boundaries are as in Figure \ref{fig:map_swc}.}
  \label{fig:map_swc_flux}
\end{figure}

Stellar density maps built from these flux chains, for the regions corresponding to Figures \ref{fig:map_ec} and \ref{fig:map_swc}, are shown in Figures \ref{fig:map_ec_flux} and \ref{fig:map_swc_flux}. For these figures, the flux chains are grouped into three metallicity populations, $-2.3 \leq [Fe/H] \leq -1.5$ (low/blue), $-1.4 \leq [Fe/H] \leq -1.0$ (intermediate/green), and $-0.9 \leq [Fe/H] \leq -0.2$ (high/red). The substructures we are interested in are most dominant in the intermediate range, so show up in the maps as a green haze. Near the East Cloud, some dwarf galaxies are clearly visible in blue, and the inner halo is also visible as a red glow to the West. The metallicity of the East Cloud will be discussed in more detail in Section \ref{sec:feh}.

The pixels are quite large compared to the maps in Section \ref{sec:density}, and thus we can only approximate the regions selected by the 25 per cent membership probability contour. The pixels selected to correspond to this region are outlined in black in both Figures \ref{fig:map_ec} and \ref{fig:map_swc}, and Figures \ref{fig:map_ec_flux} and \ref{fig:map_swc_flux}.

\subsection{Total Brightness}\label{sec:mags}

Both methods for estimating the brightness convert the CFHT $g$ and $i$ band photometry, in which the PAndAS data were taken and the flux chains were generated, into the $V$ band using the following colour transform \citep{Ibata2007}:
\[
 V_0 = \left\{ 
  \begin{array}{l l}
    -0.033 + 0.714(g_0-i_0) + i_0, & \quad (g_0-i_0)<1.850,\\
    -0.480 + 0.956(g_0-i_0) + i_0, & \quad (g_0-i_0)\geq1.850,
  \end{array} \right. 
\]

The first method is an improvement upon the method used in Paper I. We call this new method `the Dwarf Method'. We define a target region within the tangent plane projection of M31 and then select all stellar entries from the PAndAS catalogue with colours and magnitudes consistent with metal-poor red giant branch populations at the distance of M31, using the same colour-magnitude selection box as in Section \ref{sec:density}. Due to the disrupted nature of the SWC and EC structures, rather than using a radial boundary, we include all stars with a membership probability $p\geq0.25$ (following Paper I). This cutoff is marked by a contour in Figures \ref{fig:map_ec} and \ref{fig:map_swc}, specifically for the target regions SWC, ECS, and ECN.

A luminosity function is built for the target from the stars within the contour. A Poisson deviate $n$ is randomly selected from the Poisson distribution with mean value $N$, where $N$ is the number of stars used to build the luminosity function. The luminosity function is then sampled $n$ times to get an initial estimate of the target flux.
Stars outside this contour are used to define the contamination for each target. A flux estimate is produced for the local contamination similarly, which is used to correct the target flux estimate. This process is repeated $10^4$ times to estimate the total flux of the target.

In order to correct for unresolved light, we measure the brightness of available Andromeda dwarf galaxies within the PAndAS footprint, and calibrate against their $V$ band brightnesses listed in \citet{McConnachie2012}. This process is covered in detail in Paper I and remains largely unchanged, still resulting in a final calibration factor of $-1.9\pm0.2$ magnitudes. The results from this method are summarized in Table \ref{tab:appmag}.

The second method, let us call this `the Population Method', uses the single stellar population chains from Martin et al (in prep) discussed in Section \ref{sec:chains}. Using the same target regions from the first method, we select which pixels to sample chains from. For a particular entry in the chains, we sum the fluxes of all the pixels in the target for each population. 

As the chains do not allow a negative contribution from any of the populations, there is a small statistical positive bias that results in slightly higher counts in pixels with small numbers of stars. In addition to this, there is a small positive bias towards very metal poor populations ($[Fe/H]\leq-2.0$) for the East Cloud fields, due to an overlapping blue population from the PAndAS Milky Way Stream \citep{Martin2014} which is not included in the smooth contamination model. However, this is of little concern as both the SWC and EC objects contain a negligible number of stars of such low metallicity (see Figure \ref{fig:metals_25} in the next section).

To ensure that the contamination has been completely removed, and account for the positive bias, we select adjacent visually empty regions around the targets to estimate the local contamination flux contribution, see Figures \ref{fig:map_ec_flux} and \ref{fig:map_swc_flux}. The results from this method are summarized in Table \ref{tab:appmag}.

\begin{table}
\centering
\caption{Apparent magnitude estimates\label{tab:appmag}}
\begin{tabular}{ c|c|c }
\hline
Target & Dwarf Method & Population Method \\
\hline
SWC & $12.6\pm0.3$ & $13.1\pm0.2$ \\ 
ECS & $13.8\pm0.4$ & $14.3\pm0.3$ \\ 
ECN & $14.3\pm0.5$ & $14.8\pm0.3$ \\ 
ECS and ECN & $13.3\pm0.4$ & $13.8\pm0.3$ \\
\hline
\end{tabular}
\end{table}

While the results from the two methods shown in Table \ref{tab:appmag} are consistent, there is a $0.5$ magnitude systematic offset between them. This indicates that the $-1.9$ magnitude correction factor used in the Dwarf Method may be too large by $0.5$ magnitudes. This offset is not entirely surprising, as this adjustment matches the SWC and EC structures to Andromeda's dwarf galaxy population, which are much less extended and disrupted. The Population Method does not require any such adjustment factor, instead using population fits to the PAndAS data directly. Assuming the weak link to be the $-1.9$ magnitude correction used in the Dwarf Method, and applying adjustments based on the distances in Section \ref{sec:distances}, gives the absolute magnitude estimates shown in Table \ref{tab:absmag}. We have attributed the $0.5$ magnitude systematic offset to the Dwarf Method correction factor, however, it is worth noting that this value could be indicative of possible systematic errors for Population Method. 

Although both methods use the PAndAS data, they approach the problem of brightness estimation from very different angles, so the fact that they produce such consistent results, not just for the East Cloud but also for the South West Cloud, bolsters confidence.

\begin{table}
\centering
\caption{Absolute magnitude estimates\label{tab:absmag}}
\begin{tabular}{ c|c }
\hline
Target & Magnitude \\
\hline
SWC & $-11.3\pm0.3$ \\
ECS & $-10.2\pm0.4$ \\
ECN & $-9.7\pm0.4$ \\
ECS and ECN & $-10.7\pm0.4$ \\ 
\hline
\end{tabular}
\end{table}

\subsection{Surface Brightness}\label{sec:surf_mags}

To calculate the surface brightness, the Dwarf Method assumes the area of the object to be that contained within the contour representing $25$ per cent of the peak stellar density, shown in Figures \ref{fig:map_ec} and \ref{fig:map_swc}. The large pixel size used in the Population Method makes surface brightness estimation difficult for such small objects, but pixels corresponding to this contour were selected manually to best approximate the region.

The results of the two methods for computing surface brightness are consistent for the SWC,
however due to the extreme faintness of the EC structures, consistent results are not recovered for the EC; the results are summarized in Table \ref{tab:surf}.
Each method is prone to different issues for such small faint structures as the EC. The Dwarf Method suffers significantly trying to estimate the surface brightness of the EC, as the contamination in this region is very strong, suggesting that a population fitter would fare much better. Unfortunately the Population Method suffers as well due to the small size of the structure. As the pixel size used in the Population Method is quite large (a necessity to get reasonable signal for the fit), it makes the estimation of the area within such a specific and small contour very inaccurate.
The true value of the surface brightness of these structures is likely to sit between the results of the two methods.
It remains that both structures have very low surface brightness, around $32$ magnitudes per square arc second.

\begin{table}
\centering
\caption{Surface brightness estimates\label{tab:surf}}
\begin{tabular}{ c | c | c } 
\hline
Target & Dwarf Method & Population Method \\
\hline
SWC & $31.8\pm0.3$ & $31.7\pm0.6$ \\ 
ECS & $31.6\pm0.4$ & $32.8\pm0.2$ \\ 
ECN & $31.6\pm0.5$ & $33.2\pm0.2$ \\ 
ECS and ECN & $31.6\pm0.4$ & $33.0\pm0.2$ \\
\hline
\end{tabular}
\end{table}

\section{Metallicities}\label{sec:feh}

\begin{figure*}
  \includegraphics[width=100mm]{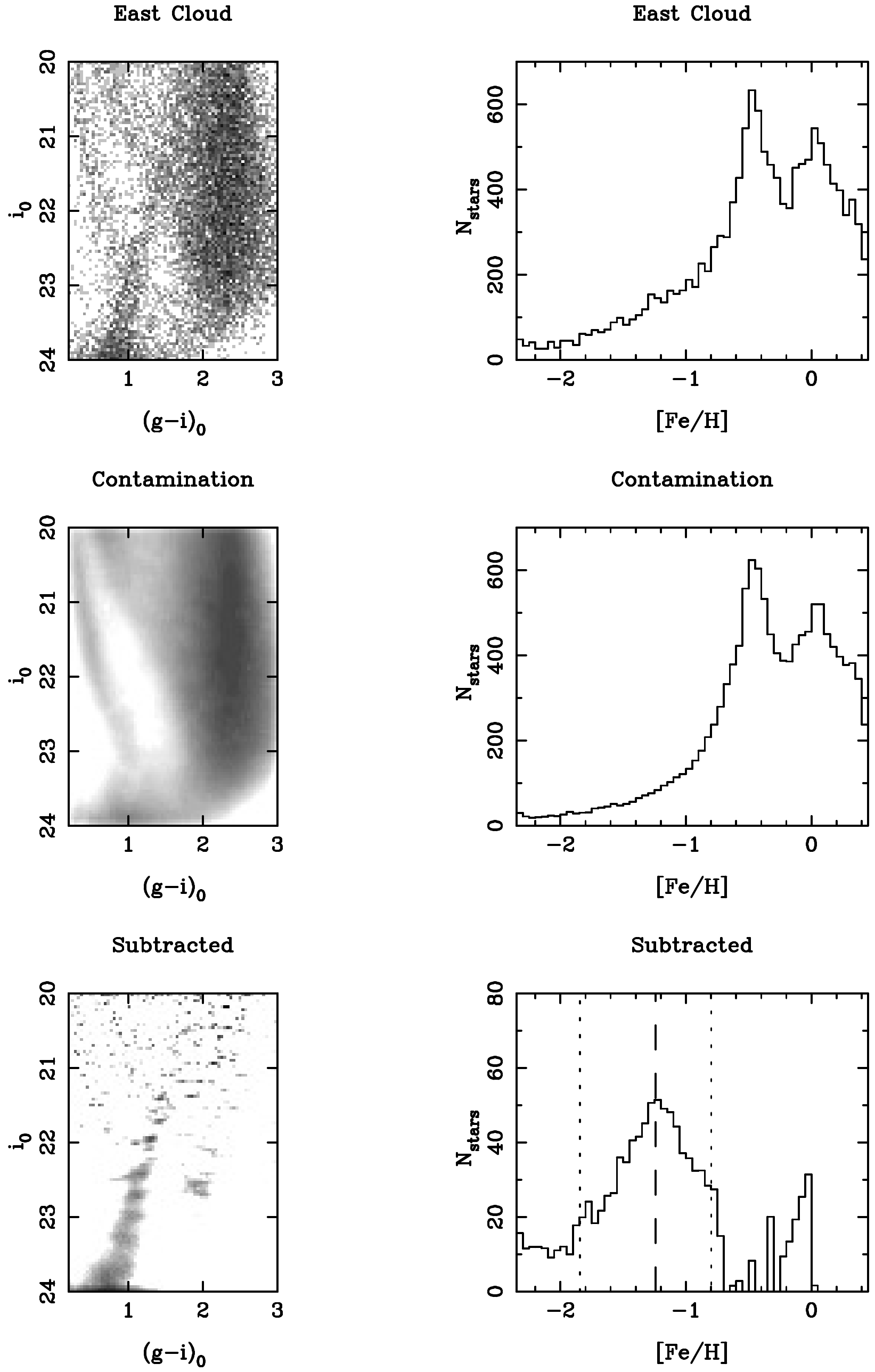}
  \caption{Left: Hess diagrams for the East Cloud (top), the contamination model for the same region (middle), and the Monte Carlo sampled residual of the East Cloud minus the contamination (bottom). The Hess diagrams are displayed with logarithmic scaling, and contain only stars with a membership probability $p\geq0.25$. Right: metallicity distribution functions generated using The Dartmouth Stellar Evolution database 2012 isochrones for a 13 Gyr population with $[\alpha/H] = 0.0$ \citep{Dotter2008}, and assuming a distance to the East Cloud of $814$~kpc. In the bottom right panel, the median is marked by a dashed line, and the dotted lines contain $68$ per cent of the distribution.}
  \label{fig:metals_25}
\end{figure*}

Using the TRGB method in Section \ref{sec:distances}, we estimated the metallicity of the East Cloud to be $-1.4^{+0.2}_{-0.1}$ dex. Here we use the technique from Paper I to estimate the metallicity, to ensure a consistent comparison between the East Cloud and the SWC.

We estimate metallicities for the stars in the East Cloud by comparison with theoretical isochrones from the Dartmouth Stellar Evolution Database \citep{Dotter2008}. We restrict ourselves to isochrones with a 13Gyr age, $[\alpha/\rmn{Fe}]=0.0$, and metallicity $-2.5 \leq [\rmn{Fe}/\rmn{H}] \leq 0.5$ in 0.05 dex steps. Dartmouth isochrones are available in CFHT/MegaCam filters, so no filter transformations are necessary.

Isochrones are shifted to the best fit East Cloud distance discussed in Section \ref{sec:distances}, 814kpc. The dereddened $g_0-i_0$ colours and $i_0$ magnitudes of each individual star are compared to the grid of isochrones, and each star is assigned the metallicity of the nearest isochrone. Stars further than 0.05 in colour or magnitude from any isochrone are excluded. Furthermore, we use only stars in regions ECS and ECN (see Figure \ref{fig:map_ec}), with membership probabilities $p\geq0.25$, as defined in Section \ref{sec:density}. 

In this way, we are able to build up metallicity distribution functions (MDFs) for East Cloud stars. In the top panels of Figure \ref{fig:metals_25}, we show the results of this process: the left panel shows a Hess diagram for stars with East Cloud membership probability $p\geq0.25$, and the right panel shows the resulting MDF. 

As we have already seen, the East Cloud is heavily affected by foreground contamination. To account for this, we use the contamination model presented in \citet{Martin2013}. As described in \citet{Bate2014}, we use this model to construct Hess diagrams and an MDF for the foreground contamination in East Cloud pixels above the membership probability cutoff. The contamination model is shown in the middle panels of Figure \ref{fig:metals_25}. We note that in order to use the contamination model, we must restrict ourselves to the following colour and magnitude ranges: $0.2 \leq (g_0-i_0) \leq 3.0$, $20 \leq i_0 \leq 24$.

The contamination-subtracted Hess diagram and MDF are shown in the bottom panels of Figure \ref{fig:metals_25}. There is a clear peak in the MDF at $[\rmn{Fe}/\rmn{H}]\sim-1.2$, which we identify as the East Cloud.

The median metallicity we measure for the East Cloud is $[\rmn{Fe}/\rmn{H}] = -1.2\pm0.1$. This measurement takes into account errors in stellar magnitudes by Monte Carlo sampling the PAndAS catalogue 1000 times, and uncertainty in the estimated East Cloud distance determined in the previous section. As expected, this measurement is consistent with the result obtained from the distance measurement technique in Section \ref{sec:distances}, which used the same isochrone set. 

As in \citet{Bate2014}, we measure the error-corrected standard deviation of the metallicity distribution, which we call the metallicity spread. This is calculated by treating photometric errors as metallicity differences $\Delta[\rmn{Fe}/\rmn{H}]$, and then Monte Carlo sampling 1000 times to build up an average $\Delta[\rmn{Fe}/\rmn{H}]$ distribution. The width of this distribution gives us the contribution of the photometric errors to the observed metallicity spread. When subtracted in quadrature, we obtain an intrinsic metallicity spread of $0.58\pm0.03$. We also repeated our analysis using tighter RGB colour-magnitude cuts (see Section \ref{sec:density}); the median metallicity is unchanged, however the metallicity spread is marginally smaller.

\section{Globular Clusters}\label{sec:gcs}

\begin{figure}
\begin{center}
\includegraphics[width=0.9\columnwidth,clip=true]{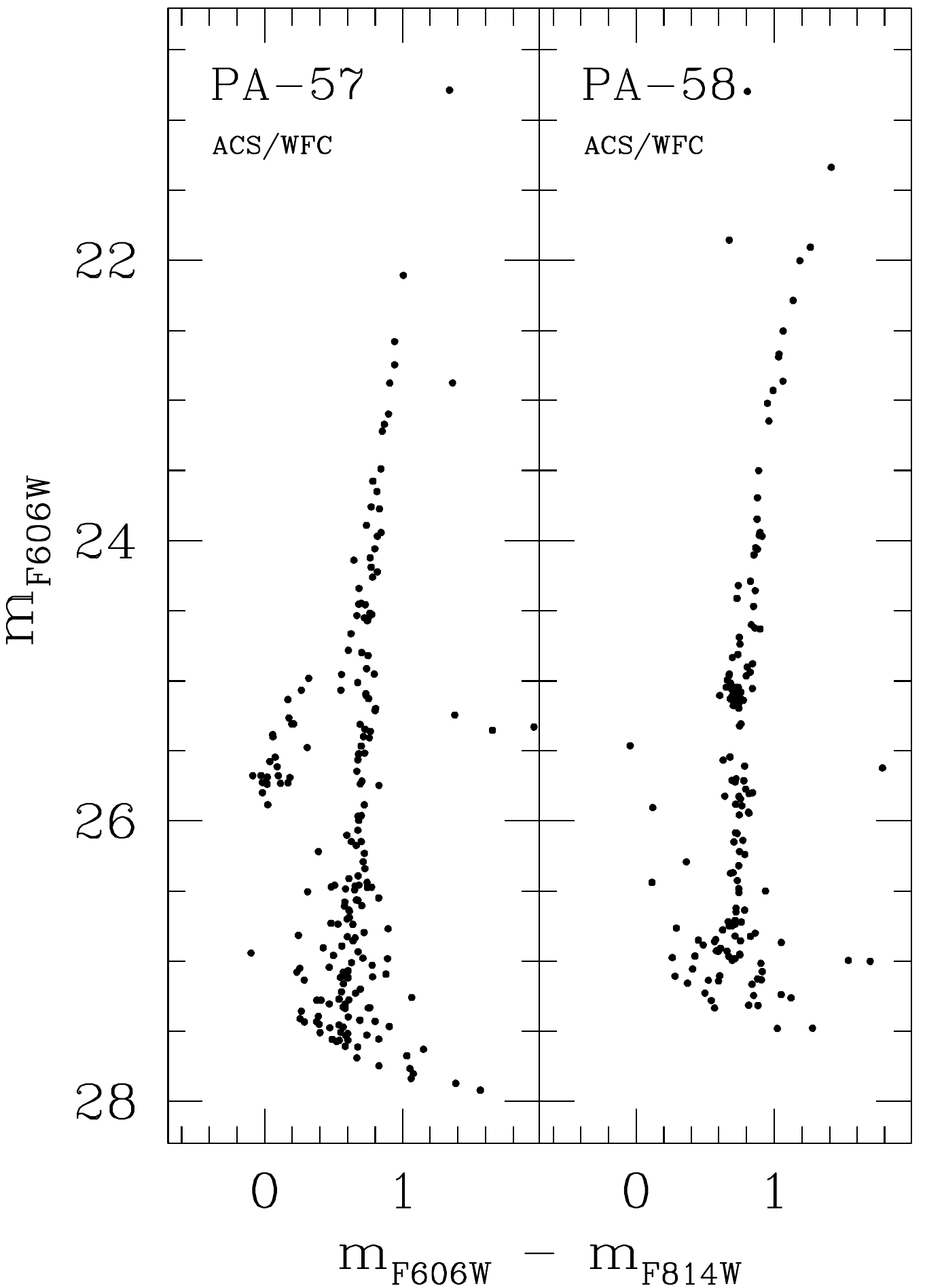}\\
\end{center}
\caption{ACS/WFC CMDs for PA-57 and PA-58.  Only stars within $30\arcsec$ of the cluster centres are plotted.
\label{f:gc_cmds}}
\end{figure}

\begin{figure}
\begin{center}
\includegraphics[width=0.9\columnwidth,clip=true]{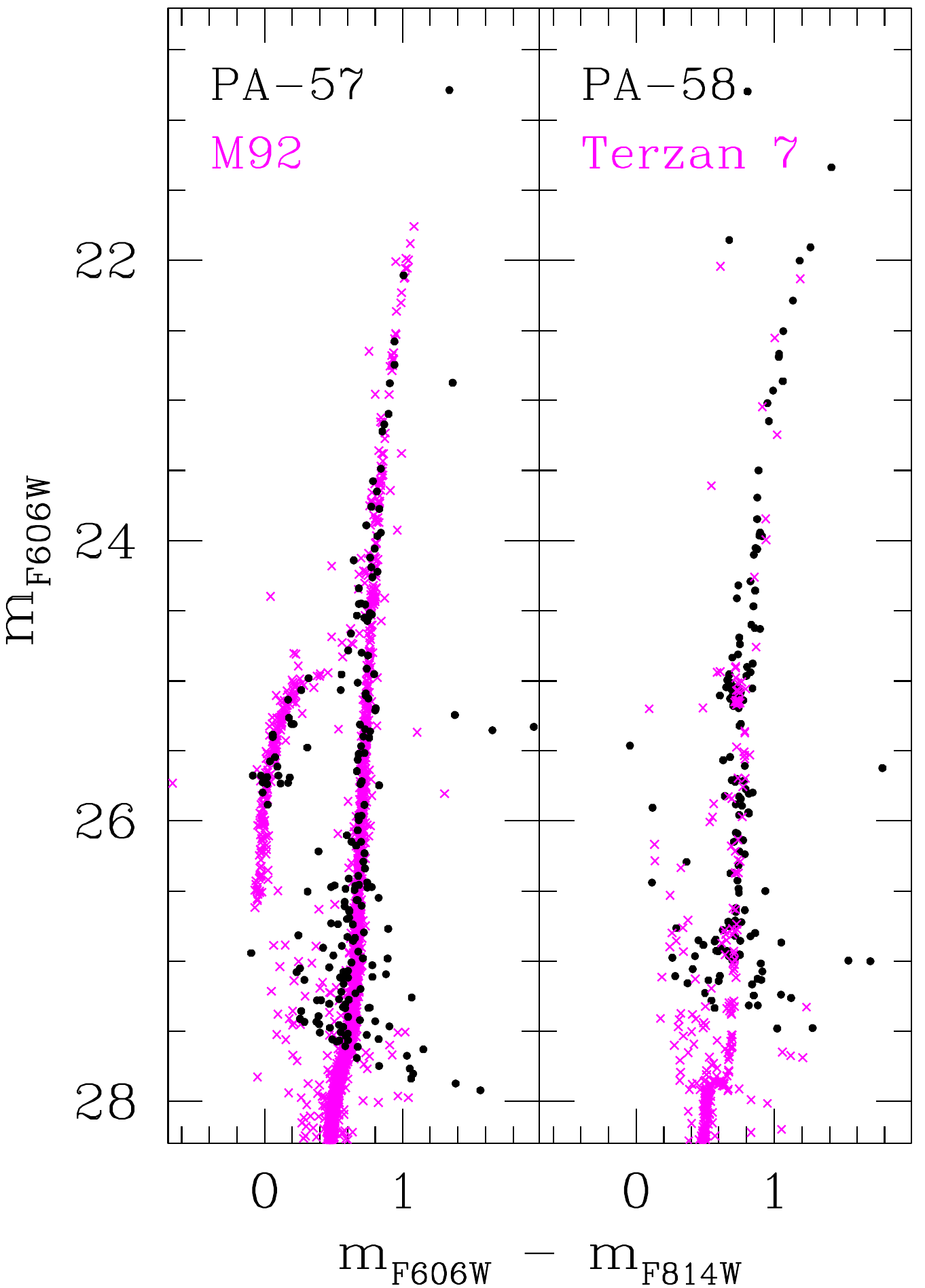}\\
\end{center}
\caption{CMDs for PA-57 and PA-58, aligned with the template clusters M92 and Terzan 7, respectively.
\label{f:gc_align}}
\end{figure}

Two M31 globular clusters (GCs) have positions that project onto the main body of the East Cloud -- these are PA-57 and PA-58 
\citep[see][]{huxor:14}. These clusters also share closely matching radial velocities that are separated from the M31 systemic velocity by 
$\sim 100$ km s$^{-1}$. The apparent spatial coincidence of PA-57, PA-58, and the East Cloud at a radius where the surface density of both GCs and substructures is extremely low, combined with the matching radial velocities of PA-57 and PA-58, can be used to argue on a statistical basis that all three are very likely to be physically associated (see \citealt{Veljanoski2014})\footnote{Of course, conclusive verification requires a velocity measurement derived directly from the East Cloud itself.}.

We obtained deep {\it Hubble Space Telescope} imaging of both PA-57 and PA-58 as part of program GO-12515 (PI: Mackey), using the
Wide Field Channel (WFC) of the Advanced Camera for Surveys (ACS). The data were taken on 2011 December 01 and 2012 March 05
for PA-57 and PA-58, respectively.  Both clusters were imaged three times with the F606W filter and three times with the F814W filter, with
small telescope dithers between exposures. Single-frame integration times were $821$s in F606W and $867$s in F814W.

Reduced ACS/WFC data obtained from the {\it HST} archive have already been corrected for charge-transfer efficiency losses using a 
pixel-based method \citep[e.g.,][]{anderson:10}. We photometered these corrected images using version $2.0$ of the {\sc dolphot} package
\citep{dolphin:00} to perform point-spread function (PSF) fitting, following a procedure very similar to that described by 
\citet{mackey:13b}.  For both clusters we set the {\sc dolphot} parameter {\tt fitsky}$=3$, to fit the sky for each star locally within the
photometry aperture as part of a two-parameter PSF fit. To excise spurious detections, non-stellar objects, and objects with poor photometry 
from our catalogues we selected only those sources classified as stellar by {\sc dolphot}, with valid measurements in all six input frames, 
and possessing a global sharpness parameter between $\pm 0.05$ in each filter, and a crowding parameter $\leq 0.07$ mag in each filter.  
The final photometry is on the VEGAMAG scale of \citet{sirianni:05}. 

We use artificial star tests to quantify the individual photometric uncertainties and assess the detection completeness. For each real
star we generate $500$ artificial stars of the same brightness, with positions uniformly distributed within an annulus spanning
$0.5\arcsec-5\arcsec$ of its coordinates. Using {\sc dolphot} we add a single artificial star to the image at a time, and then
attempt to measure it. Once this process is complete we apply the same quality filter as described above. For each real star we 
then assign a detection completeness according to the ratio of the number of successfully measured artificial stars to the number
of input stars, and obtain an estimate of the photometric uncertainty by quantifying the scatter in the differences between input 
and measured magnitudes for the artificial stars. In uncrowded parts of the target clusters (i.e., outside $\approx 5\arcsec$ from
their centres) the $50\%$ completeness level is at $m_{\rm F606W} \approx 27.2$ and $m_{\rm F814W} \approx 26.6$; however in the
central regions these limits are as much as $\sim 1.5$ mag brighter. 

\begin{table*}
\centering
\caption{Results of CMD registration between our target and template clusters.}
\begin{tabular}{@{}lcccccccccc}
\hline \hline
Name & $[$Fe$/$H$]$ & \hspace{1mm} & $\Delta c$ (mag) & $\Delta m$ (mag) & \hspace{1mm} & $E(B-V)_{\rm template}$ & $\mu_{\rm template}$ & \hspace{1mm} & $E(B-V)_{\rm cluster}$ & $\mu_{\rm cluster}$ \vspace{1mm}\\
\hline
{\bf PA-57} & & & & & & & & & $(0.066)$ & \vspace{1mm}\\
M92 & $-2.4$ & & $0.050\pm0.005$ & $8.25\pm0.05$ & & $0.021$ & $16.43$ & & $0.074$ & $24.55$ \\
M53 & $-2.0$ & & $0.045\pm0.005$ & $9.95\pm0.05$ & & $0.023$ & $14.80$ & & $0.071$ & $24.63$ \vspace{2mm}\\
{\bf PA-58} & & & & & & & & & $(0.062)$ & \vspace{1mm}\\
Terzan 7 & $-0.6$ & & $-0.030\pm0.005$ & $7.50\pm0.05$ & & $0.088$ & $17.02$ & & $0.056$ & $24.60$ \\
Palomar 12 & $-0.8$ & & $0.045\pm0.005$ & $8.27\pm0.05$ & & $0.037$ & $16.40$ & & $0.085$ & $24.55$ \\
\hline
\label{t:gc_align}
\end{tabular}
\end{table*}

Figure \ref{f:gc_cmds} shows our colour-magnitude diagrams (CMDs) for PA-57 and PA-58. To minimise contamination from 
non-cluster members we plot only stars lying within $30\arcsec$ of the cluster centres. The two CMDs have quite
distinct morphologies indicating that the clusters have rather different properties. PA-57 has a steep red giant branch (RGB) and
a blue horizontal branch (HB), characteristic of an ancient ($> 10$ Gyr old), metal-poor population. PA-58, on the other hand,
has an upper RGB that is substantially more curved, and a strikingly short HB that is exclusively red -- in fact, more of a red
clump than a typical globular cluster HB.  In this sense it strongly resembles two of the clusters known to be associated with
the South-West Cloud -- PA-7 and PA-8 -- which have inferred ages at least $2$ Gyr younger than the oldest Milky Way 
globular clusters \citep[see][]{mackey:13a}.

We estimate the metallicity, foreground reddening, and line-of-sight distance of PA-57 and PA-58 by matching fiducial 
sequences from several Milky Way globular clusters to their CMDs. Appropriate template clusters are selected from the
sample observed by the ACS Galactic globular cluster Treasury Survey \citep[see][]{sarajedini:07}, and photometry was obtained
from the online database for this project\footnote{\url{http://www.astro.ufl.edu/~ata/public_hstgc/}} \citep{anderson:08}.
In identifying suitable template clusters, we search for objects possessing comparable upper RGB curvature and HB morphology
to PA-57 and PA-58. A small amount of experimentation revealed the clusters M92 (NGC 6341) and M53 (NGC 5024) to be
good matches for PA-57, and the clusters Terzan 7 and Palomar 12 to be good matches for PA-58.

We align the template CMDs to those of PA-57 and PA-58 using a procedure very similar to that described by 
\citet{mackey:06,mackey:07}. The vertical offset, $\Delta m$, required to register the two CMDs was determined largely via the
observed HB level, while the horizontal offset, $\Delta c$, is determined using the colour of the RGB near the HB level.
Adopting suitable extinction coefficients from \citet{schlafly:11} -- i.e., $A_{\rm F606W} = 2.471 E(B-V)$ and 
$A_{\rm F814W} = 1.526 E(B-V)$, where $E(B-V)$ is determined from the maps of \citet{Schlegel1998} -- allows the foreground
reddening and distance modulus for a given target cluster to be expressed in terms of the same quantities for the
template clusters, and $\Delta m$ and $\Delta c$:
\begin{align}
E(B-V)_{\rm cluster}&= 1.058\Delta c + E(B-V)_{\rm template}\nonumber\\[6pt]
\mu_{\rm cluster}&= \Delta m - 2.615\Delta c + \mu_{\rm template}\,.
\end{align}
Thus the final uncertainty on measurements for the target cluster is determined by both the precision to which the
CMDs can be registered, and the precision with which the reddening and distance of the template object are known.
Note that this assumes the target cluster and the template cluster are identical in terms of age, $[$Fe$/$H$]$, and 
$[\alpha/$Fe$]$; small variations in these properties result in additional second-order systematic uncertainties.

The results of our alignment process are shown in Table \ref{t:gc_align}, and examples of registered CMDs are displayed in Figure
\ref{f:gc_align}. The assumed foreground reddening for the template clusters comes from the \citet{schlegel:98} maps; also listed 
(in parentheses) are the values derived from these maps for the target clusters, which can be used as a consistency check on the 
measurements determined using $\Delta c$. Values for the metallicity and distance of the template clusters come from 
\citet{dotter:10}, who derived these quantities directly from the same photometric data set we employ here. The one 
exception is the distance for Terzan 7, which we take from \citet{siegel:11} (who again used the Treasury survey photometry, 
but studied Terzan 7 in detail). Uncertainties on the template distances are difficult to ascertain -- \citet{dotter:10} do not provide 
formal uncertainties on their measurements, but there are substantial variations amongst the literature -- see for example the 
2010 revision of the \citet{harris:96} catalogue. Terzan 7 represents an extreme example -- \citet{dotter:10} find a distance modulus 
of $17.15$ for this cluster, while the Harris catalogue lists $16.79$, derived from the horizontal branch luminosity measured by \citet{Buonanno1995}. Our assumed value represents a suitable compromise.
Given the excellent precision with which we are able to register the CMDs, it is safe to assume that uncertainties in the known 
distances to the template clusters dominate our overall error budget, such that the uncertainties on our individual distance 
measurements are of order $\sim 0.1$ mag.

Overall, we find that PA-57 is a very metal-poor cluster, with $[$Fe$/$H$]$ somewhere between $-2.0$ and $-2.4$ -- its CMD 
matches those for M92 and M53 equally well. PA-58 is much more metal-rich, with $[$Fe$/$H$] \sim -0.7$.  Given that both 
Terzan 7 and Palomar 12 are substantially younger than the oldest Galactic GCs (\citet{dotter:10} find ages for these clusters of
$8.00\pm 0.75$ Gyr and $9.50\pm 0.75$ Gyr, respectively), the good agreement between the CMD for PA-58 and those for the
template clusters suggests that PA-58 is likely to be a similarly young cluster. These results are strongly suggestive that the 
progenitor of the East Cloud underwent significant chemical evolution over many Gyr before it was destroyed by M31.

We have determined two independent distance measurements to the East Cloud by matching two different templates to each of PA-57 and PA-58. A straight average of the results gives a distance modulus of $24.58$ with an uncertainty of
order $\sim 0.05$ mag. This corresponds to a line-of-sight distance of $824\pm19$ kpc.

\section{Discussion and Conclusions}\label{sec:discussion}

Like the SWC covered in Paper I, the East Cloud is one of the brightest stellar substructures in the outer halo of M31 today. 
We can use the luminosity-metallicity relation \citep{Kirby2011} to estimate the luminosity of the East Cloud progenitor:
\begin{equation}
[Fe/H] = -2.06 + (0.40 \pm 0.05) \rmn{log_{10}}\left(\frac{L_{\rmn{tot}}}{10^5L_\odot}\right)
\end{equation}
For a median metallicity of $[\rmn{Fe}/\rmn{H}] = -1.2$, the progenitor luminosity is $\sim1.4\times10^7L_\odot$, which corresponds to an absolute V-band magnitude of $M_V \approx -13.0$~mag. 
This high metallicity implies that the East Cloud progenitor was of comparable brightness to the SWC progenitor, and thus was amongst the brightest of Andromeda's satellites. However, the East Cloud is much more extended and diffuse than the SWC, making characterisation more difficult. 

Comparing with the brightness estimate determined in Section \ref{sec:brightness}, the luminosity-metallicity relation suggests that the present day East Cloud represents approximately $12$ per cent of the progenitor's original luminosity. With clear signs of disruption, it is likely that we are seeing a heavily disrupted remnant today. 
Key properties of the East Cloud, including distances, luminosities, and metallicities have each been calculated using multiple methods, and are summarised in Table \ref{tab:properties}. 

\begin{table}
\centering
\caption{Properties of the East Cloud\label{tab:properties}}
\begin{tabular}{ l c } 
\hline
Milky Way distance & $814^{+20}_{-9}$~kpc\\
M31 distance & $111^{+13}_{-1}$~kpc\\
Apparent $V$-band magnitude & $13.8\pm0.3$~mag\\
Absolute $V$-band magnitude & $-10.7\pm0.4$~mag\\
Average $V$-band surface brightness & $31.6\pm0.4$~mags/arcsec$^2$\\
Metallicity $[Fe/H]$ & $-1.2\pm0.1$\\
Metallicity spread $\Delta[Fe/H]$ & $0.58\pm0.03$\\
\hline
\end{tabular}
\end{table}

Distances to the main East Cloud substructure were calculated using a TRGB method, detailed in another paper in this series (Conn et al submitted). 
By fitting templates to the coincident globular clusters PA-57 and PA-58, we established an independent measure of the distance to the cloud, under the assumption that the clusters are associated. 
Comparing stars in the East Cloud to theoretical stellar isochrones for a $13$~Gyr population, gives a metallicity of $[Fe/H] = -1.2\pm0.1$. This is consistent with the metallicity parameter returned by TRGB method fit. 
Metallicities were also calculated for the two globular clusters coincident with the East Cloud. Given the substantial difference between their metallicities, we suspect the East Cloud progenitor underwent significant chemical evolution prior to its eventual disruption. This is consistent with the broad MDF found for the East Cloud substructure.

The luminosity was estimated using two methods, the `Dwarf Method' and the `Population Method'. The `Dwarf Method' is calibrated against estimates of the luminosities of the Andromeda dwarf galaxies. The `Population Method' is directly estimated from forward modelled single stellar population fits to the PAndAS data. These methods returned consistent estimates for the SWC and the East Cloud, and are an improvement on the work presented in Paper I.

With the new lower estimate of $-11.3\pm0.3$~mag for the absolute magnitude of the SWC, we can say that it is much more disrupted than previously thought, with the current visible structure representing only approximately $40$ per cent of the luminosity of the progenitor object estimated by Paper I. 
It is unclear where to draw the boundaries for such extended objects as the SWC and the EC. As discussed in Paper I, there is a significant extension to the South-East of the SWC which may potentially be part of the SWC, and could represent a portion of the absent $60$ per cent of the the progenitor. 

It is even less clear where to draw the boundaries for the East Cloud. ECS and ECN are disjointed, but could easily be part of the same progenitor, as we have assumed throughout, and further to this there are signs of more potentially related structure in ECNW to the West of ECN and in ECSW to the South-West of ECS. Adding these structures to the total for the East Cloud would raise the total brightness by a further half a magnitude, and would suggest a much more diffuse structure overall. 

However, even if we include the extended wings, this would only raise the portion of the implied progenitor that we are observing from $12$ to $18$ per cent. It is clear that the East Cloud is much more disrupted than the SWC. 
This conclusion is supported by its lower surface brightness and greater physical extent. Furthermore, unlike the SWC, there are no signs of nearby HI \citep{Lewis2013}, suggesting the gas was stripped at an earlier stage, and that the East Cloud has either been disrupted more violently or over a longer timescale than the SWC.

The fact that two GCs are seen to be associated with the core of the East Cloud, and possibly up to another four with its full extent, adds further weight to the argument that we are only observing a relatively small fraction of the original progenitor.  A useful (albeit crude) statistic in this context is the globular cluster specific frequency, which is defined as the number of GCs per unit luminosity of the host galaxy, normalised to a host with $M_V = -15$. Specifically, $S_N = N_{GC} 10^{0.4(M_V + 15)}$, where $N_{GC}$ is the total number of globular clusters \citep{Harris1981}.

Assuming our derived luminosity for the core of the East Cloud, plus the association of PA-57 and PA-58, returns a specific frequency of $\sim 166$. This is much larger than is commonly seen for other dwarf galaxies (\citealt{Miller1998}; \citealt{Miller2007}; \citealt{Georgiev2010}; \citealt{Harris2013}). If we instead use our estimate of the progenitor luminosity using the metallicity derived in Section \ref{sec:feh}, we find $S_N \sim 13$, which sits right in the midst of the typically observed range for dwarfs. Similarly, the present luminosity of the SWC together with its three coincident GCs suggests $S_N \sim 91$, which is very high; however, using our estimate of the progenitor luminosity derived from the median metallicity reduces this to $S_N \sim 33$.

It is interesting to speculate briefly as to whether the four GCs projected onto the two features extending away from the East Cloud might also be associated with this structure. Including all four would raise the specific frequency to $\sim 38$, which is not unreasonably high; furthermore, the velocities for these clusters measured by \citet{Veljanoski2014} exhibit a gradient that is not inconsistent with a stream-like association. HST photometry for these objects would enable precise distance measurements that would more definitively test the possibility of their membership in the underlying substructure, and help place the locus of the substructure in three dimensions relative to the M31 centre. Metallicity information for the GCs could also provide information about the chemical evolution of the East Cloud progenitor.

Ideally, follow up spectroscopy will confirm the associations between the extension fields ECSW and ECNW, the core fields ECS and ECN, and the globular clusters, but this is likely to be extremely challenging due to the very high levels of contamination in the extensions, combined with the very diffuse substructure. 
Outer halo substructures as faint and diffuse as the East Cloud represent the limit of what is possible to recover from the PAndAS data alone.

\section*{Acknowledgments}

We wish to thank the anonymous reviewer for helpful suggestions
which significantly improved this paper. 
NFB and GFL thank the Australian Research Council (ARC) for support through Discovery Project (DP110100678). GFL also gratefully acknowledges financial support through his ARC Future Fellowship (FT100100268). BM acknowledges the support of an Australian Postgraduate Award. ADM is supported by an Australian Research Fellowship from the ARC (DP1093431). 
This work is partially based on observations made with the NASA/ESA \textit{Hubble Space Telescope}, obtained at the Space Telescope Science Institute (STScI), which is operated by the Association of Universities for Research in Astronomy, Inc., under NASA contract NAS 5-26555. These observations are associated with program GO 12515. 

\bibliography{Substructure,gc_text}

\bsp

\label{lastpage}

\end{document}